# Robust pinned magnetisation in $A_2$Ir$_2$O$_7$ iridates, the case of Er$_2$Ir$_2$O$_7$ and Lu$_2$Ir$_2$O$_7$ flux-grown single crystals

Daniel Staško[1,*], Filip Hájek[1], Kristina Vlášková[1], Jiří Kaštil[2], Margarida Henriques[2], Milan Klicpera[1]

[1]Charles University, Faculty of Mathematics and Physics, Department of Condensed Matter Physics, Ke Karlovu 5, 12116 Prague 2, Czech Republic

[2]Institute of Physics of the Czech Academy of Sciences, Na Slovance 1999/2, 182 21 Prague 8, Czech Republic

*Corresponding authors: daniel.stasko@matfyz.cuni.cz

**Abstract:**

Reliable and profound studies of actual magnetic domain structure in rare-earth A$_2$Ir$_2$O$_7$ pyrochlore iridates are frequently limited by insufficient sample quality or lack of single crystals. We report the magnetic properties of the for-the-first-time synthesised Lu$_2$Ir$_2$O$_7$ and Er$_2$Ir$_2$O$_7$ single-crystals. The paper is focused on the robust ferromagnetic component of magnetisation present in the material with the antiferromagnetically ordered state of the all-in-all-out (AIAO) type, and is discussed in the framework of AIAO and AOAI domains and interfaces on the geometrically frustrated lattice.

Systems incorporating both spin-orbit coupling (SOC) and electron correlations, such as the members of the $A_2B_2O_7$ family with $A$ = rare-earth element and $B$ = $4d$ or $5d$ transition metal, present a fertile area for research of novel exotic phases of matter [1,2,3]. Relativistic SOC by itself plays an essential role in the emergence of non-trivial topology in the band structures of these materials [3,4]. In particular, the $A_2Ir_2O_7$ iridates make for an intriguing case by combining comparable strength of SOC and electron correlations, as well as *f-d* exchange coupling between rare-earth and Ir moments, and crystal field effects. Several non-trivial phases have been proposed to arise from these interactions, including antiferromagnetic Weyl semimetal [5,6,7], topological Mott insulator [7,8], axion insulator with non-collinear magnetic order [5,9], spin-ice state with monopole-like excitations [10,11], fragmented state [12,13], or spin-liquid state [14].

Complex electronic and magnetic states in $A_2Ir_2O_7$ are also closely connected with their geometrically frustrated crystal lattice of the pyrochlore type. Pyrochlore structure is composed of two sublattices of $A^{3+}$ and $Ir^{4+}$ ions, forming interpenetrating networks of corner-sharing tetrahedra and eight- and six-coordinated oxygen cages around respective cations [1,15]. Focusing exclusively on the Ir sublattice, it has been repeatedly proposed and validated that it orders magnetically in the so-called all-in-all-out (AIAO and equi-energy AOAI) structure when the magnetic moments on vertices of the Ir tetrahedron point all in or out along local <111> directions [16,17]. With the exception of $Pr_2Ir_2O_7$ without any signs of magnetic ordering [18], heavier $A_2Ir_2O_7$ display the magnetic transition at 30-140 K depending on the $A$ ion [1,3,15]. Simultaneously with the magnetic ordering, a metal-to-insulator transition is observed [19, 20], and proposed to be induced by the antiferromagnetic ordering of the Ir sublattice [21,22]. Importantly, both the AIAO and energy-equivalent AOAI arrangements of magnetic moments could coexist in the lattice, forming the respective domains [1,2,3]. The interfaces between domains, AIAO/AOAI domain walls (DWs), seem to exhibit significantly different conductive properties compared to the interior of the domain [23]. It was demonstrated that insulating domains have relatively conductive domain walls with disturbed magnetic ordering. The perturbed ordering in the walls leads to uncompensated magnetic moments on the interfaces [24,25], or even magnetic monopole-like excitations in the rare-earth sublattice induced by the *f-d* coupling [11].

In the present paper, the magnetic response of the Ir sublattice to the applied magnetic field is discussed within the model of antiferromagnetic (AFM) domains and robust ferromagnetic (FM) domain walls in two representatives of the $A_2Ir_2O_7$ family: $Lu_2Ir_2O_7$ and $Er_2Ir_2O_7$ single crystals. Solely Ir magnetism rules over the magnetic states in $Lu_2Ir_2O_7$ with nonmagnetic $Lu^{3+}$ ions, representing the model system for studying the Ir sublattice. $Er_2Ir_2O_7$ properties are, in addition, influenced by the magnetic $Er^{3+}$ sublattice, especially at low temperatures.

Single crystals of $Lu_2Ir_2O_7$ and $Er_2Ir_2O_7$ were synthesised for the first time, characterised by electron microscopy (Fig. 1e-f), X-ray diffraction methods, and investigated by means of magnetisation measurements. $PbF_2$-flux method was used for sample synthesis, resulting in minor Pb contamination; details on sample synthesis and characterisation are provided in Supplementary materials. Both single crystals reveal a bifurcation between zero-field-cooled (ZFC) and field-cooled (FC) magnetic susceptibility (Fig. 1a-b), tracking the magnetic ordering of the $Ir^{4+}$ sublattice. The ordering temperature $T_{Ir}$ is determined as an onset of the bifurcation: $T_{Ir}$ = 128(3) K for $Lu_2Ir_2O_7$ and $T_{Ir}$ = 120(6) K for $Er_2Ir_2O_7$. Previous studies of polycrystalline samples reported somewhat higher ordering temperatures of 140 K [26,27,28] for $Er_2Ir_2O_7$, and 147 K [29] and 135 K [30] for $Lu_2Ir_2O_7$. The value of $T_{Ir}$ in $A_2Ir_2O_7$ iridates is strongly sample-dependent and can be connected with an Ir off-stoichiometry [24,31], which is a common problem in synthesising these materials [30,32,33,34]. Smaller $T_{Ir}$ of our single crystals is likely related to a slight Ir off-stoichiometry of up to 2% (minor Pb contamination of the samples should not affect the Ir sublattice substantially); see details in Supplementary materials. Nevertheless, such a small Ir off-stoichiometry is considered to impact the physical properties of the material only slightly ($T_{Ir}$ values close to the polycrystalline ones) compared to most of the previously reported $A_2Ir_2O_7$ crystals.

Focusing on the magnetic field dependence of magnetisation (Fig. 1c-d), the magnetisation does not saturate up to 7 T, nevertheless a clear saturation tendency is followed. Roughly one-third of the free-ion saturated magnetisation (1 $\mu_B$ for $Ir^{4+}$ and 9 $\mu_B$ for $Er^{3+}$ ions) is reached in 7 T. Further increase of magnetisation with increasing field is expected, likely reaching one-half of the free-ion value in the magnetic field well above 10 T. Such an evolution is perfectly in agreement with previous studies of both $A_2Ir_2O_7$ single crystals and polycrystals [26,35,11]; being explained by strong geometrical frustration of the pyrochlore lattice and local magnetic moments anisotropy. The average magnetic anisotropy of the compound, measuring the magnetisation with a magnetic field applied along three principal crystallographic directions, is relatively low (Fig. 1c-d). The largest magnetisation is observed for a field applied along the [110] direction and the smallest one along the [111] direction in both $Lu_2Ir_2O_7$ and $Er_2Ir_2O_7$. The fact that the same anisotropy is revealed by both members strongly suggests only a minimum impact of the Er sublattice, which is expected to order in an easy-plane arrangement [26]. Indeed, an anomaly connected with magnetic correlations on the Er sublattice was reported at a much lower temperature of 0.6 K [26]. On the other hand, the impact of the Ir sublattice on the Er sublattice is demonstrated by a more substantial anisotropy (magnetisation in 7 T differs for [110] and [111] directions by 0.3 $\mu_B$ in $Lu_2Ir_2O_7$ and 1.0 $\mu_B$ in $Er_2Ir_2O_7$), likely reinforced by Ir-polarised Er moments. A crude explanation of the magnetic anisotropy in the system is based on the orientation of sublattice tetrahedra with respect to an applied magnetic field. The AIAO ordering of magnetic moments is characterised by the moments oriented parallel or antiparallel to local <111> directions. When a magnetic field is applied along this direction, it is naturally difficult to polarise the moments due to their antiferromagnetic exchange coupling on the geometrically frustrated tetrahedron. When a field is applied along another crystallographic direction, the system robustness to a field polarisation is lower as the vector of the magnetic field is not parallel to any direction of moments on the tetrahedra. No metamagnetic transition is observed in the data in a field up to 7 T. A standard temperature evolution of isothermal magnetisation in an antiferromagnet is followed (insets in Fig. 1c-d).

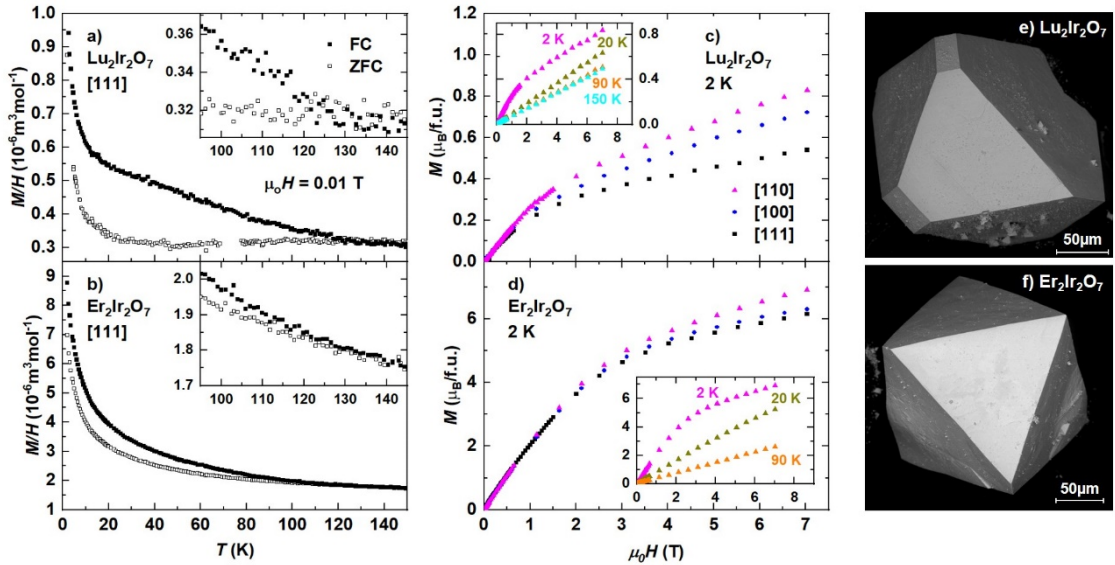

Fig. 1: Characterisation of $Lu_2Ir_2O_7$ (top panels) and $Er_2Ir_2O_7$ (bottom panels) single crystals. a-b) Bifurcation between zero-field-cooled and field-cooled magnetic susceptibility with the zoomed-in high-temperature region in the insets (field applied along the [111] direction). c-d) Field dependence of magnetisation at 2 K with field applied along the three principal crystallographic directions.

Magnetisation along [110] was measured at several temperatures (insets). e-f) Back-scattered electron (BSE) images of the synthesised single crystals.

Ferromagnetic ordering in the investigated crystals is excluded based on the missing hysteresis loops measuring the magnetisation with an applied field of 0 → 7 → -7 → 0.2 T. However, a clear FM response is revealed (Fig. 2a-b) by measuring the samples with the following cooling protocol: before cooling the sample below the ordering temperature $T_{Ir}$, an external magnetic field is applied. The sample is cooled in this field (FC) down to low temperatures, and the field is removed. The additional magnetic signal induced by this cooling protocol is represented by a small shift of magnetisation $M_{sh}$ (insets of Fig. 2a-b). Importantly, this signal is present in any magnetic field, at least up to 7 T (-7 T); $M_{sh}$ is robust against the applied field (Fig. S1). No shift is observed in the ZFC regime. $M_{sh}$ is positive or negative depending on the sign of an applied cooling field and is symmetric around the ZFC value. That is, $M_{sh}$ measured in 7 T–FC is equal to $-M_{sh}$ measured in -7 T–FC. Such a FM contribution is ascribed to the disturbed magnetic ordering on the antiferromagnetic AIAO–AOAI interfaces, in agreement with some light rare-earth $A_2Ir_2O_7$ [24,36,37] and $Cd_2Os_2O_7$ [25]. The domain wall model is characterised by the pinned magnetic moments on the interfaces between the AIAO and AOAI domains. The pinned moment is created upon the formation of the DW at $T_{Ir}$, resulting from uncompensated moments on the Ir tetrahedron (Fig. 3). One can describe this magnetic ordering as two-in-two-out (2I2O) or three-in-one-out (3I1O; alternatively 1I3O). In the ZFC regime, the DWs and related uncompensated moments are created randomly upon crossing the $T_{Ir}$. Therefore, the total net magnetisation is zero. The application of a magnetic field at low temperatures induces a standard response of an antiferromagnetic material without any field hysteresis of magnetic response. However, when a magnetic field is applied upon cooling (FC regime), crossing $T_{Ir}$, the domain walls are created in a way that has uncompensated magnetic moments oriented, preferably along the field direction. Once this ferromagnetic-like ordering on the DWs is formed, it is very robust against external magnetic field below $T_{Ir}$. The uncompensated moments on the interface are protected by the antiferromagnetic ordering of the AIAO and AOAI domains. Unless the antiferromagnetic ordering is disturbed, e.g. by increasing temperature above $T_{Ir}$, the pinned moments are extremely stable.

Investigating magnetic moments of the DWs in detail, several measurement protocols are followed: (i) field dependence of magnetisation on FC sample introduced above. Measuring the isothermal magnetisation at different temperatures, a temperature dependence of $M_{sh}$ is revealed (open symbols in Fig. 2c-d). $M_{sh}$ decreases with increasing temperature when approaching the ordering temperature $T_{Ir}$. We highlight that the $M_{sh}$ value remains constant on the field interval (-7, 7) T at any given temperature in $Lu_2Ir_2O_7$. The same behaviour is observed in $Er_2Ir_2O_7$, however, only at higher temperatures (≥ 20 K). At 2 K, the $M_{sh}$ is strongly influenced by the exchange interactions of the Er sublattice and develops with the applied magnetic field (Fig. S1). By increasing (decreasing) the field to 7 T (-7 T), $M_{sh}$ decreases, approaching zero in high field. Such an evolution can be attributed to the stronger exchange interactions between Er moments and their *f-d* coupling with Ir moments, including the pinned moments, in high magnetic field. Nonetheless, the reversibility of the processes must be emphasised; after removing the field, the $M_{sh}$ shift changes to its original value. The same evolution of $M_{sh}$ is observed for the magnetic field applied along all three principal crystallographic directions.

(ii) temperature dependence of magnetisation in zero field after field-cooling (the measurement scheme is introduced in the inset of Fig. 2 d). That is, the ferromagnetic component induced in DWs by the FC protocol is followed (Fig. 2c-d). Importantly, the measured magnetisation corresponds to $M_{sh}$ and its temperature dependence follows the values of $M_{sh}$ determined from the field measurements (Fig. 2a-b). A similar value of $M_{sh}$ observed in $Lu_2Ir_2O_7$ was previously reported for $Eu_2Ir_2O_7$ single crystal [24] and $Lu_2Ir_2O_7$ polycrystals [30]. Only small anisotropy in $M_{sh}(T)$ is observed, in contrast to the isothermal magnetisation measurements in Fig.1c-d. The most striking difference between $Lu_2Ir_2O_7$ and $Er_2Ir_2O_7$ members is observed at the lowest temperatures, where $M_{sh}$ is strongly enhanced in $Er_2Ir_2O_7$.

The impact of the Ir molecular field and exchange interactions between Er moments are proposed to play a significant role at these temperatures. Notably, $M_{sh}$ is dependent on the cooling field; higher magnetic field leads to higher values of $M_{sh}$ (Fig. 2e).

(iii) a cooling-field evolution of a bifurcation between ZFC and FC magnetisation (Fig. 1a-b and Fig. S2). Subtracting the ZFC from the FC magnetisation (denoted as $\Delta M$ in Fig. 2e-f) reveals temperature evolution almost identical to the evolution of $M_{sh}$ discussed in the previous paragraph (see 0.01 T measurements in Fig. 2e). $\Delta M$ increases with increasing cooling field for both studied compounds. Similar values of $\Delta M$ and $M_{sh}$, their temperature and field evolution and respective curve shapes strongly suggest that the bifurcation between ZFC and FC magnetisation below $T_{Ir}$ is directly connected with the FM component $M_{sh}$ described by the domain interfaces in these materials. Increasing the cooling field (FC) results in an increase of $M_{sh}$, suggesting a larger density (larger number or size/thickness) of DWs.

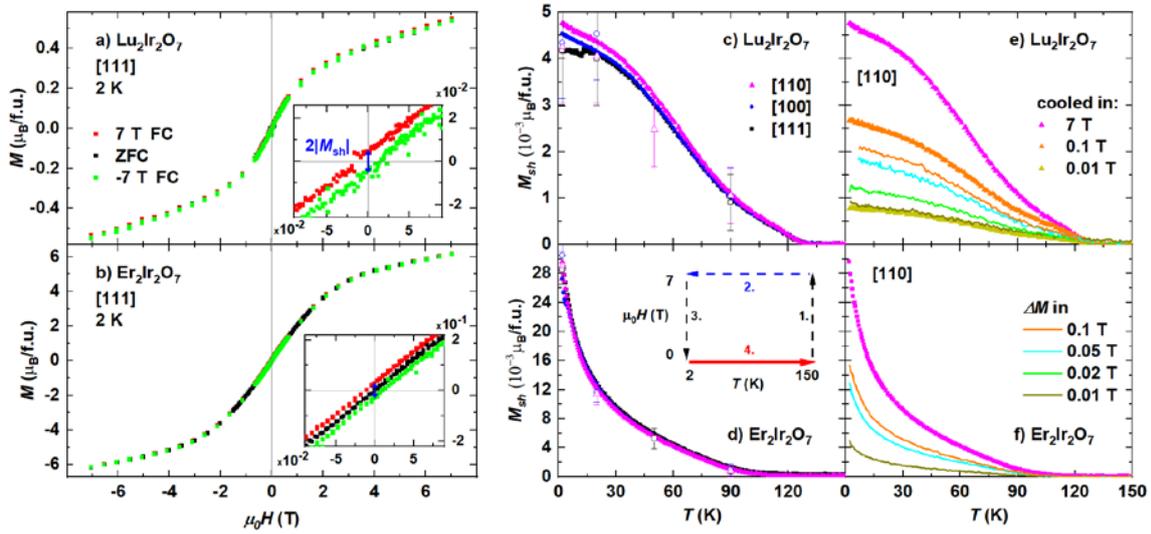

Fig. 2: Ferromagnetic signal in $Lu_2Ir_2O_7$ (top panels) and $Er_2Ir_2O_7$ (bottom panels) single crystals. a-b) Field hysteresis measured in different field-cooled protocols. No hysteresis is observed in the ZFC or FC data; however, a clear FM contribution to the magnetisation $M_{sh}$ accompanies the 7 T-cooled and -7 T-cooled data (see also the insets). c-d) Temperature dependence of $M_{sh}$ measured along the three principal crystallographic directions in 0 T after cooling down in 7 T (see the scheme in panel b). Open symbols indicate values determined from the field measurements in panels a) and b). e-f) Temperature dependence of $M_{sh}$ ([110] direction displayed) is compared with $\Delta M$ determined as the difference between ZFC and FC magnetisation in Fig. 1a-b and Fig. S2 (same units and scale for $\Delta M$ and $M_{sh}$ are used).

Domain wall structure (simple examples introduced in Fig. 3) is supposed to be strongly dependent on the direction of the applied magnetic field. In the presented examples, the net magnetic moment of the domains' interface is aligned along the external field and perpendicular to the DW. Of course, more complex DWs with net magnetisation not necessarily perpendicular to the DW also exist. Nevertheless, these simple DW models can be sufficiently employed for rudimentary explanations of our experimental results. First, the total net magnetisation per one DW in one unit cell is calculated. The same value of 14.4% of the saturated moment is computed for the three types of DWs (see details of calculations in Supplementary materials). That is, in the theoretical case of a perfect crystal with the same amount of DWs of a single type, $M_{sh}$ would be identical for the three principal crystallographic directions. Such estimation is in agreement with only very weak anisotropy of experimental $M_{sh}$ (Fig. 2c-d).

Secondly, the dimensions of the AFM domains are roughly estimated in $Lu_2Ir_2O_7$, assuming only the $Ir^{4+}$ ions' magnetism. Considering the first type of DWs in Fig. 3a, the minimum distance between individual DWs along the net moment direction is estimated to be 0.064 µm (that is, 253 Ir layers) based on the measured value of $M_{sh}$ in 7 T-FC regimen (see Supplementary materials for calculation details). Counting in also DWs parallel to the net magnetisation, simple cubic domains with dimensions of 0.19x0.19x0.19 µm³ are estimated. Previously, AFM domains were directly observed in $Nd_2Ir_2O_7$ [23] and $Cd_2Os_2O_7$ [38], with dimensions of microns and dozens of microns in ZFC, respectively. Importantly, the application of magnetic field during sample cooling below $T_{Ir}$ reduced the number of domains, all the way to a single-domain state in the high-field-cooled regimes [23,39]. Such an observation contrasts with the increasing value of $M_{sh}$ with increasing cooling field (our work and similarly previous works [24,25,30]) – increasing value of $M_{sh}$ is explained by a higher concentration of DWs and, therefore, a higher number of domains. Following this explanation, the density of domains in $Lu_2Ir_2O_7$ single crystal is calculated to be 146 AIAO/AOAI domain interfaces per µm³. Alternatively, increasing thickness of DWs (simply stacking more layers of tetrahedra on top of each other) can result in the same net magnetisation with bigger domain sizes compared to the presented monolayer model. Stacking an odd number of layers creates a conventional DW, while stacking an even number of layers creates a boundary between the same type of domains (two AIAO or two AOAI). That is, a single domain would possess the pinned moments. Of course, FM droplet model could also be considered: the FM component $M_{sh}$ would be attributed to Ir sublattice defects caused by off-stoichiometry [30,40]. However, previous studies of iridates' off-stoichiometry showed a gradual decrease in net magnetisation with increasing off-stoichiometry [24,31,41], disagreeing with the FM droplet model.

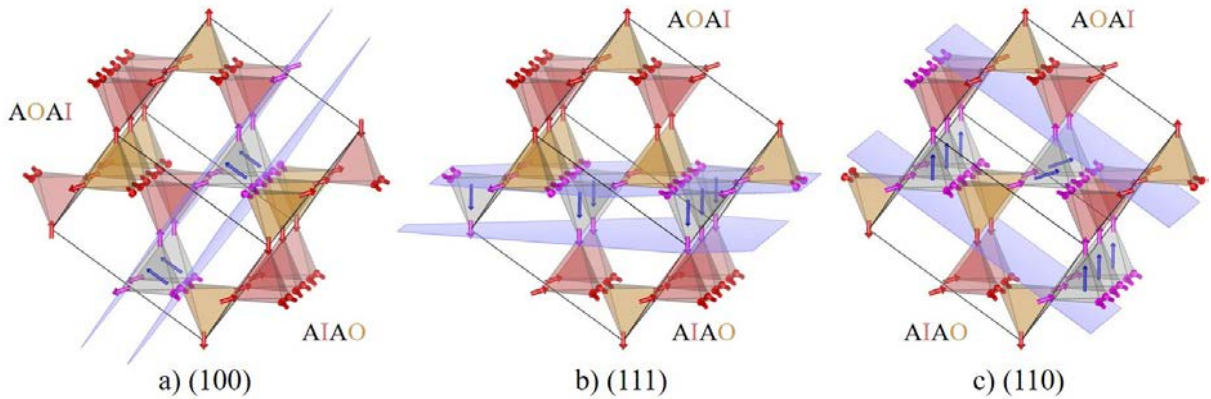

Fig. 3: Domain walls between AIAO and AOAI domains in the pyrochlore structure (cubic unit cell is outlined) induced by a magnetic field applied along the three principal crystallographic directions. The tetrahedra with compensated magnetic moments, all-in (red) or all-out (yellow), and uncompensated moments (grey) within the domain wall are depicted. The moments on the borders of the domain wall are drawn as pink arrows, in contrast to red arrows in the interior of domains. Total pinned magnetic moment (blue arrow) points perpendicular to the domain wall. a) 2-in-2-out case, pinned moment along [100], b) 3-in-1-out case, pinned moment along [111] and c) combined 3-in-1-out and 1-in-3-out case to create the pinned moment along [110].

In conclusion, the magnetic response of two representatives of $A_2Ir_2O_7$ iridates, $Lu_2Ir_2O_7$ and $Er_2Ir_2O_7$ single crystals, was investigated and discussed within the framework of antiferromagnetic domains and ferromagnetic domain interfaces model. Ferromagnetic component of magnetisation was shown to manifest as (i) a shift in the value of isothermal magnetisation measured on the FC sample, (ii) a temperature evolution of magnetisation of the FC sample measured in zero magnetic field, and (iii) very importantly a bifurcation between ZFC and FC magnetisation below $T_{Ir}$. The three evolutions give the same value of the ferromagnetic component, proving the same origin of respective signals. Simultaneously, it was proved that the ferromagnetic component in $A_2Ir_2O_7$ can be effectively investigated by any of the three methods. A model of pinned magnetic moments in domain walls between antiferromagnetic domains was employed to describe the measured ferromagnetic signal. The model allowed us to explain a bifurcation in magnetisation of antiferromagnetically AIAO ordered iridates, the negligible anisotropy of the ferromagnetic response, as well as to estimate the size and number of the antiferromagnetic domains in the studied single crystals. Robustness of the domain walls below ordering temperature was proved.

**Acknowledgements:**

Sample synthesis and characterisation were done in MGML (mgml.eu), which is supported within the program of Czech Research Infrastructures (project no. LM2023065). The study was supported by Charles University, project GA UK (no. 148622), and Barrande Mobility project (no. 8J24FR013).

## Supplementary materials

**Single crystal synthesis and characterisation:**

Single crystals of $Lu_2Ir_2O_7$ and $Er_2Ir_2O_7$ iridates were synthesised employing a lead-fluoride flux preparation method. The mixture of initial $Lu_2O_3$ ($Er_2O_3$), $IrO_2$ and $PbF_2$ was properly homogenised, inserted into a platinum crucible and reacted at 1100 °C in air in a resistance furnace. Further details on the synthesis process and its optimisation will be reported in our separate publication. Grown single crystals have an octahedral (bi-pyramidal) shape (Fig. 1e-f) with pronounced shiny facets perpendicular to the <111> crystallographic directions and with dimensions up to 0.5x0.5x0.5 mm$^3$. The basic characterisation was done on multiple crystals from different batches. Laue X-ray diffraction confirmed the single-grain nature of larger crystals and was used for sample orientation. Electron microscopy with energy-dispersive X-ray (EDX) analysis revealed the stoichiometry of investigated crystals. Besides Lu (Er), Ir and O atoms, the EDX spectra indicated the presence of lead in the samples under the surface. Lead atoms from the $PbF_2$ flux tended to mix with the rare-earth atoms in the pyrochlore structure. This was evident from a homogenous distribution of Ir in the 2D EDX maps while regions with more Pb showed smaller signal from Lu/Er and vice versa. Overall, a small doping of lead was observed (up to 10% of rare-earth atoms). However, some crystals showed up to approximately 40% of rare-earth atoms being substituted by lead. Those crystals were excluded from further investigations. Within the experimental error of about 4%, an almost stoichiometric ratio of Ir and Lu(Er)+Pb was confirmed, the ratio of 0.98(4):1.02(4). The oxygen composition could not be reasonably determined as the EDX method is not sufficiently sensitive to lighter atoms.

The crystal structure was confirmed to be of the pyrochlore-type (cubic ordered structure, space group *Fd-3m*) by single-crystal diffraction on $Lu_2Ir_2O_7$ and $Er_2Ir_2O_7$ samples of dimensions 0.13×0.07×0.04 mm$^3$ and 0.13×0.10×0.03 mm$^3$, respectively. The crystals were glued on top of a glass fibre and mounted onto a goniometer head. The diffracted intensities were collected at 120 K using a four-circle diffractometer (Gemini from Rigaku Oxford Diffraction) equipped with a Mo x-ray tube [$\lambda(MoK_\alpha)$ = 0.71073 Å], Mo-enhance collimator, graphite monochromator and an Atlas CCD detector. The CrysAlisPro [42] software was used to collect and reduce the data, and to face-index the crystal shape, necessary to perform the empirical absorption correction using spherical harmonic functions implemented in the SCALE3 ABSPACK scaling algorithm. The Superflip [43] program was used for structure solution and the Jana2020 package [44] for structure refinement based on $F^2$. The refinement confirmed the correctness of the structural model with the *Fd-3m* space group with *R*-factors converging to $R_{obs}$ = 5.1% and 5.5% for $Lu_2Ir_2O_7$ and $Er_2Ir_2O_7$, respectively. The lattice parameters obtained from the single-crystal x-ray diffraction are *a* = 10.1215(1) Å for $Lu_2Ir_2O_7$ and *a* = 10.1626(1) Å for $Er_2Ir_2O_7$. The atomic occupancy factors were first refined and later fixed as there were no significant deviations from full occupancy. Mixed occupancies for Lu/Er and Ir with Pb were also tried, leading to worse refinement factors compared to the sole occupation of the respective positions. The refined atomic positions and equivalent isotropic displacement parameters are given in Table S1.

Table S1. Atomic coordinates ($x$, $y$, $z$), equivalent isotropic displacement parameters ($U_{eq}$), and their estimated standard deviations for $Lu_2Ir_2O_7$ and $Er_2Ir_2O_7$.

| Atom | Wyckoff position | $x$ | $y$ | $z$ | $U_{eq}$(Å$^2$) |
|---|---|---|---|---|---|
| Lu | 16$d$ | 0.5 | 0.5 | 0.5 | 0.0019(2) |
| Ir | 16$c$ | 0 | 0 | 0 | 0.0021(3) |
| O1 | 8$b$ | 0.375 | 0.375 | 0.375 | 0.056(3) |
| O2 | 48$f$ | 0.338(6) | 0.125 | 0.125 | 0.023(6) |
| Er | 16$d$ | 0.5 | 0.5 | 0.5 | 0.0038(1) |
| Ir | 16$c$ | 0 | 0 | 0 | 0.0045(1) |
| O1 | 8$b$ | 0.375 | 0.375 | 0.375 | 0.011(2) |
| O2$_E$ | 48$f$ | 0.332(3) | 0.125 | 0.125 | 0.045(4) |

**Magnetisation measurements**

Magnetisation measurements were performed using an MPMS7-XL instrument (Quantum Design), which employs a high-resolution superconducting quantum interference device (SQUID). The reciprocating sample option (RSO) was used to measure relatively weak magnetic signal from small single crystals. All measurements were done on the same sample of $Lu_2Ir_2O_7$ and $Er_2Ir_2O_7$, with a mass of 0.14 mg and 0.09 mg, respectively. The shape of the crystals and related demagnetisation factor were neglected. All relevant data were measured up to 150 K only. At higher temperatures (> 200 K), the magnetic signal of samples, especially $Lu_2Ir_2O_7$, became unreliable. The signal of the sample environment (diamagnetic and paramagnetic signal of plastic holders and GE varnish) contributed significantly to the measured magnetisation.

The isothermal magnetisation data were measured with different cooling protocols (ZFC, 7 T–FC and -7 T–FC); the example is presented in Fig. 2a-b in the main text. The difference between 7 T–FC and -7 T–FC datasets is followed in any given magnetic field and temperature. The magnetisation difference represents the pinned FM moment 2|$M_{sh}$|, as $M_{sh}$ is the difference between ±7 T–FC and ZFC data. Measurements at two different temperatures and with magnetic field applied along the three principal crystallographic directions are presented in Fig. S1. The $M_{sh}$ value is constant in varying magnetic field applied along the three directions in $Lu_2Ir_2O_7$, signifying the robustness of the pinned moment. The $M_{sh}$ value is slightly lower at 20 K compared to 2 K. The same development is observed for $Er_2Ir_2O_7$ at temperature ≥ 20 K. However, a decrease of $M_{sh}$ with increasing field (to 7 T or -7 T) is tracked at 2 K. This low-temperature evolution of $M_{sh}$ is attributed to stronger exchange interactions between Er moments and their coupling with the Ir sublattice.

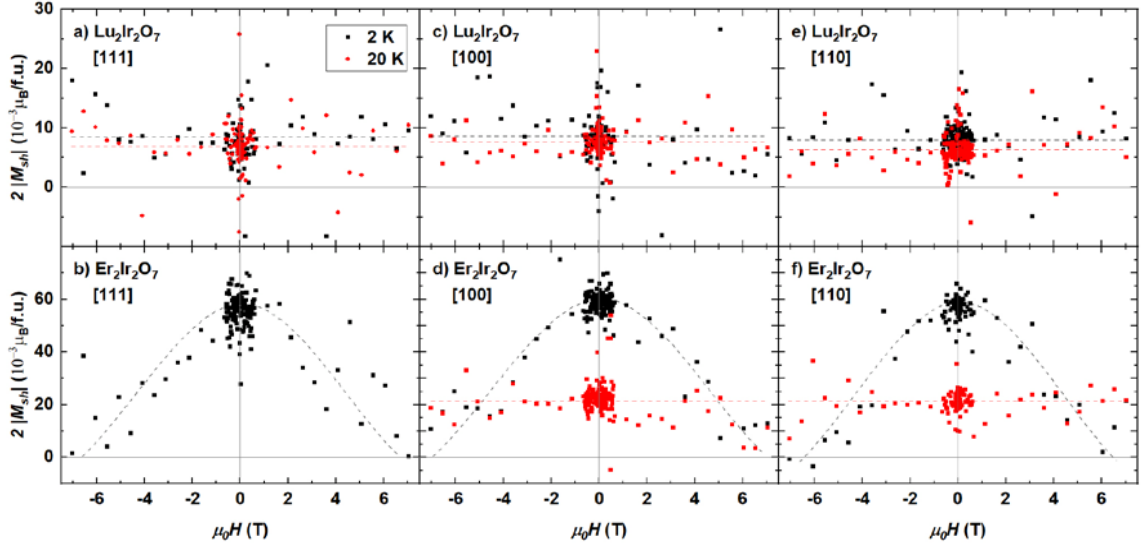

Fig. S1: Field dependence of difference data calculated by subtracting 7 T–FC and -7 T–FC datasets. The difference data represents a ferromagnetic signal of domain walls, that is, the value $2M_{sh}$ discussed in the main text. Measurements at two temperatures are presented, demonstrating an impact of the Er sublattice magnetism on the lowest temperature data of $Er_2Ir_2O_7$. Dashed lines are guides to the eye.

The temperature dependence of magnetisation under ZFC and FC regimes was measured in several magnetic fields applied along the three principal crystallographic directions (Fig. S2). Generally, the same effect of the cooling protocol was observed for all fields and directions in both $Lu_2Ir_2O_7$ and $Er_2Ir_2O_7$. Nevertheless, the difference between ZFC and FC magnetisation increased with the increasing cooling field. This increase was parametrised by subtracting the ZFC data from FC data as $\Delta M$. Temperature dependence of $\Delta M$ is presented in Fig. 2e-f in the main text. Notably, in the plots of magnetic susceptibility (one example in Fig. 1a-b), the bifurcation between ZFC and FC seems to get smaller with increasing cooling field, even though the bifurcation in magnetisation increases.

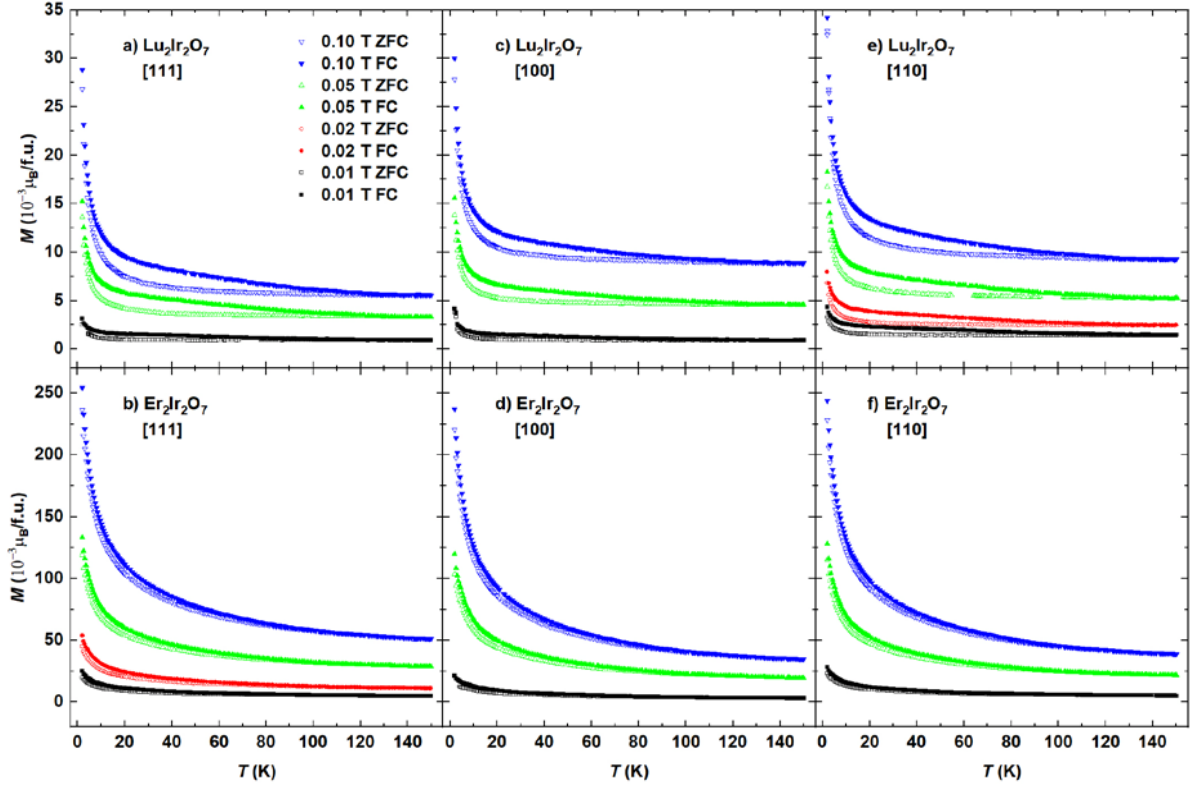

Fig. S2: Temperature dependence of magnetisation. Bifurcation between zero-field-cooled (ZFC) and field-cooled (FC) magnetisation is followed in Lu$_2$Ir$_2$O$_7$ and Er$_2$Ir$_2$O$_7$ single crystals along the three principal crystallographic directions. The data was used to calculate $\Delta M$ in Fig. 2e-f.

In addition to the measurement of individual single crystals, sets of unaligned crystals were investigated up to room temperature. Multiple crystals with a total mass of 3.43 mg for Er$_2$Ir$_2$O$_7$ and 7.68 mg for Lu$_2$Ir$_2$O$_7$ were glued together by the GE varnish in a pseudo-polycrystalline arrangement. Such samples allowed effective investigation in the whole temperature range of 2 – 350 K. The same (considering powder-like averaging) behaviour was observed on these sets of samples and single crystals reported in the main text. Moreover, the high-temperature data was analysed by fitting it to the Curie-Weiss law (Fig. S3). A strong magnetic response of the Er$_2$Ir$_2$O$_7$ sample allowed to fit the data reliably (only the high-temperature region was used for fitting; red line in Fig. S3), resulting in paramagnetic Weiss temperature $\theta_p$ = -36.4 K and effective magnetic moment $\mu_{eff}$ = 9.0 $\mu_B$. Fitted effective moment, smaller than the moments of the Er$^{3+}$ and Ir$^{4+}$ free-ions (9.58 $\mu_B$ and 1.73 $\mu_B$, respectively), illustrated a certain substitution of Er by Pb, in agreement with the EDX analysis. Our previous study of a CsCl-flux-grown polycrystalline sample revealed $\theta_p$ = -26 K and $\mu_{eff}$ = 9.12 $\mu_B$ [15,27], demonstrating by comparison a reasonably small Er off-stoichiometry of our single crystals. The much smaller magnetisation of the Lu$_2$Ir$_2$O$_7$ sample prevented us from deriving credible parameters. The diamagnetic and paramagnetic contributions of the sample environment were comparable to the sample Ir response at the highest temperatures.

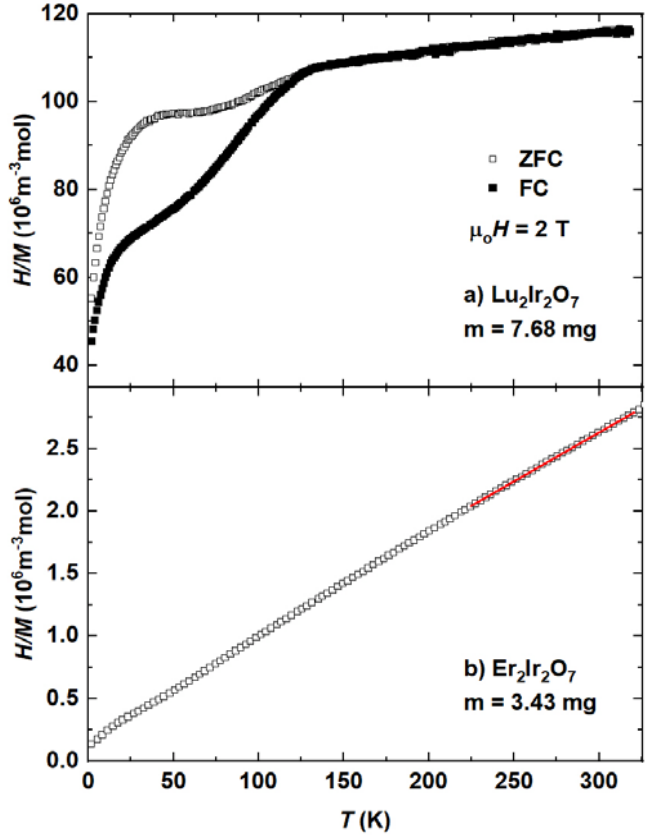

Fig. S3: Inverse magnetic susceptibility measured on sets of single crystals of a) $Lu_2Ir_2O_7$ and b) $Er_2Ir_2O_7$ glued together in a random orientation. Trials to fit the data to the Curie-Weiss law failed in $Lu_2Ir_2O_7$; both ZFC and FC magnetisation are presented. The Curie-Weiss fit to the $Er_2Ir_2O_7$ data was done on the high-temperature data (red line), resulting in parameters listed in the text.

**Determination of the total pinned magnetic moment in different types of domain walls**

We present a detailed explanation of the claim that all three main types of DWs illustrated in Fig. 3 in the main text produce an equal total magnetic moment. A single unit cell of the pyrochlore lattice contains 16 $Ir^{4+}$ cations forming 8 tetrahedra. For simplicity, an $Ir^{4+}$ moment will be used as a unit in the calculation. Each Ir cation contributes a moment of 0.5 to each of the two tetrahedra it forms. Therefore, each Ir tetrahedron, if fully polarised, hosts a maximum moment of 2. The total uncompensated magnetic moment produced by one DW with an area of $a^2$ (lattice parameter) is computed as $M_{hkl} = n_{hkl} m_{hkl}$, where $n_{hkl}$ is the number (concentration) of tetrahedra in the (hkl) plane DW with the area of $a^2$ and $m_{hkl}$ is the magnetic moment of one tetrahedron along the [hkl] direction.

(i) DW in (100) plane. The (100) case represents the simplest DW out of the three DWs presented in Fig. 3a. The width of the DW is $d_{100} = a/4$, and it is characterised by 2 out of 8 tetrahedra in the unit cell. Therefore, 2 tetrahedra ($n_{100} = 2$) form the DW per the area of $a^2$. Each of these DW tetrahedra has the 2-in-2-out order. All four spins on the tetrahedron hold the same angle with the [100] direction; that is, a projected value of the moment $\frac{1}{\sqrt{3}}$, which gives a moment $m_{100} = \frac{1}{\sqrt{3}} 4 \frac{1}{2} = \frac{2}{\sqrt{3}}$. The total moment per area $a^2$ is then $M_{100} = 2 \frac{2}{\sqrt{3}} = \frac{4}{\sqrt{3}}$.

(ii) DW in (111) plane. The (111) case is also realised with the uncompensated tetrahedra in a single layer (Fig. 3b). However, the four moments are arranged in the three-in-one-out order. The width of the DW in (111) is $d_{111} = \frac{\sqrt{3}}{6}a$; the DW in (111) case is thicker and has a higher concentration of tetrahedra per area than in (100) DW. Nevertheless, the concentration of tetrahedra in a volume is constant. That is, the concentration per area is proportional to the width of the DW, leading to the concentration $n_{111} = \frac{d_{111}}{d_{100}} n_{100} = \frac{4}{\sqrt{3}}$. One moment in a single tetrahedron is aligned along the [111] direction with a maximum contribution of 0.5, while the three remaining spins have a projection equal to 1/6. Together, they give $m_{111} = 1$ and therefore $M_{111} = \frac{4}{\sqrt{3}} 1 = \frac{4}{\sqrt{3}}$.

(iii) DW in (110) plane. The (110) case is more complex than the previous two. One (110) DW consists of two layers of tetrahedra (Fig. 3c), but only half of the tetrahedra inside the DW are uncompensated. For better clarity, the calculation is done for one layer composed of only uncompensated tetrahedra. Such monolayer has a width of $d_{110} = \sqrt{2}a/4$, which is the largest out of the three cases and results in the largest concentration of $n_{110} = \frac{d_{110}}{d_{100}} n_{100} = 2\sqrt{2}$. Similarly to the case of (111) DW, there is a 3-in-1-out (or 1-in-3-out) order. However, this order can result in an uncompensated moment along only <111> directions. Therefore, an uncompensated moment of half of the tetrahedra is pointing along a different direction than for the second half, which compensates into a moment along [110]. This moment is reduced to a projection of $m_{110} = \frac{\sqrt{2}}{\sqrt{3}} m_{111} = \frac{\sqrt{2}}{\sqrt{3}}$ for every tetrahedron, resulting in the total moment $M_{110} = 2\sqrt{2} \frac{\sqrt{2}}{\sqrt{3}} = \frac{4}{\sqrt{3}}$.

The three DW models presented in Fig. 3, corresponding to the simple types of domain walls between AIAO and AOAI domains induced by a magnetic field applied along the three principal crystallographic directions, result in the same value of the total uncompensated moment. Such a result agrees with our experimental observation of $M_{sh}$ and its minimal anisotropy discussed in the main text.

**Estimation of the domains' dimensions**

With the knowledge of the total ferromagnetic moment from a single DW, a very rough estimation of the domains' dimensions can be made based on our experimental data. Considering a domain structure with only DWs perpendicular to the [100] direction (Fig. 3a) and one DW intersecting each unit cell, the ferromagnetic moment of each unit cell is exactly $\frac{4}{\sqrt{3}}$ out of the maximum value of 16. That is, approximately 14.4% of the maximum moment is expected.

Based on the temperature evolution of the ferromagnetic component in Lu$_2$Ir$_2$O$_7$ (Fig. 2c), the $M_{sh}$ value would reach a value of approximately 0.0023 $\mu_B$/Ir$^{4+}$ at temperature approaching 0 K. The magnetic moment of a single Ir is estimated to be 1 $\mu_B$ (free ion model). $M_{sh}$ value thus represents approximately 0.23% of the saturated moment.

Employing the simple model introduced above, approximately every 63$^{rd}$ unit cell is required to have a DW to reach the measured value of the $M_{sh}$ component. Using the lattice parameter $a = 10.1215$ Å, the width of the domains (distance between individual DWs along the [100] direction) reaches 0.064 μm. Of course, this value is just a rough estimate of the domain width, considering only one type of domain. Counting in also DWs which are parallel to the net uncompensated magnetisation would lead to larger cubic domains of 0.19x0.19x0.19 μm³.